\documentclass[12pt]{iopart}
\usepackage{iopams}
\usepackage{graphicx}
\usepackage{color}
\usepackage{ulem}

\newcommand{\text}{\mathit}

\begin{document}
\title[Spherically symmetric Riemannian manifolds of constant scalar curvature\dots]
{Spherically symmetric Riemannian manifolds of constant scalar curvature and their conformally flat representations}

\author{Patryk Mach$^1$ and Niall \'O~Murchadha$^2$}
\address{$^1$M.~Smoluchowski Institute of Physics, Jagiellonian University, Reymonta 4, 30-059 Krak\'{o}w, Poland}
\address{$^2$Physics Department, University College Cork, Cork, Ireland}
\ead{\mailto{Patryk.Mach@uj.edu.pl}; \mailto{niall@ucc.ie}}

\begin{abstract}
All spherically symmetric Riemannian metrics of constant scalar curvature in any dimension can be written down in a simple form using areal coordinates. All spherical metrics are conformally flat, so we search for the conformally flat representations of these geometries. We find all solutions for the conformal factor in 3, 4 and 6 dimensions. We write them in closed form, either in terms of elliptic or elementary functions. We are particularly interested in 3-dimensional spaces because of the link to General Relativity. In particular, all 3-dimensional constant negative scalar curvature spherical manifolds can be embedded as constant mean curvature surfaces in appropriate Schwarzschild solutions. Our approach, although not the simplest one, is linked to the Lichnerowicz--York method of finding initial data for Einstein equations. 
\end{abstract}


\section{Introduction}

Spherical geometries with constant scalar curvature appear in many guises in solutions to the Einstein equations. If the scalar curvature is negative, they are the simplest constant mean curvature (CMC) slices through the Schwarzschild solution --- the umbilical ones. They also appear as moment of time symmetry slices through the Schwarzschild--anti-de Sitter spacetime. If the scalar curvature is positive, they appear as the standard closed universes in the Friedmann--Lema\^{\i}tre--Robertson--Walker cosmologies. They are also the simplest slices through the Schwarzschild--de Sitter spacetime.

Because all spherical geometries are conformally flat, it is interesting to search for conformally flat representations of spherically symmetric metrics of constant scalar curvature. While, of course, we are particularly interested in the case of 3 dimensions, in principle this task makes sense in any dimension. The equation for the conformal factor is deceptively simple. In $N \geq 3$ dimensions it has the form
\begin{equation}
\label{Phi}
\nabla^2\phi \pm \phi^{(N+2)/(N-2)} = 0,
\end{equation}
with the plus sign for positive scalar curvature and the minus sign for negative, and where $\nabla^2$ represents the flat Laplacian on $\mathbb R^N$. The 2-dimensional case can be considered separately, but it is not of much interest to our discussion.

In this paper, we want to investigate the above problem in all dimensions $N \geq 3$ and for both signs of the scalar curvature. While many special solutions of Eq.~(\ref{Phi}) exist, we can only find general analytic solutions when $(N + 2)/(N -2)$ is an integer. This is only true for $N = 3$, 4 and 6. We focus on these three cases and find closed form expressions for all spherical solutions of Eq.~(\ref{Phi}). In general they are given in terms of Jacobi or Weierstrass elliptic functions, but for some special cases it is possible to write down the solutions using elementary functions.

Because of the links to General Relativity and the Einstein constraints, we focus most our attention on the negative scalar curvature case with $N = 3$. This is done in Sections 2 and 3. In Section 2 we describe in detail the umbilical constant mean curvature slices through both the positive and negative mass Schwarzschild solutions. These represent the geometries we are interested it. In Section 3 we study the solutions of Eq.~(\ref{Phi}) with negative sign and $N = 3$ and write down all solutions in closed form. We show the relationship between these geometries and the umbilical slices through the Schwarzschild spacetime.

In subsequent sections we deal with higher dimensional cases. In particular, we list all solutions in dimensions $N = 4$ and $N = 6$.

\section{Constant negative scalar curvature spherical 3-metrics in the Schwarzschild solution}

\begin{figure}
\includegraphics{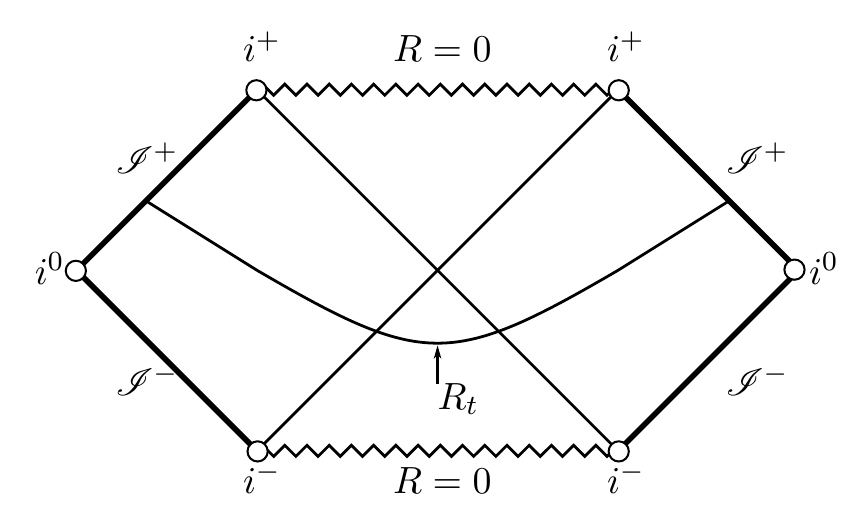} \includegraphics{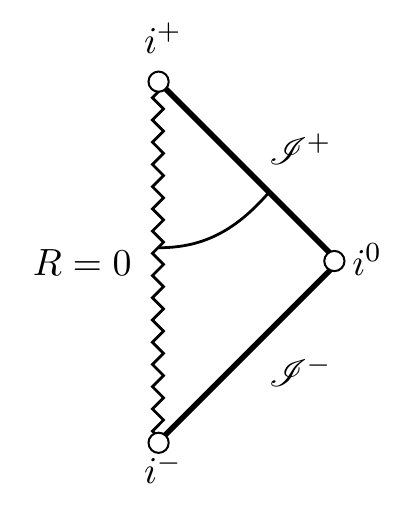}
\caption{\label{fig13}Embeddings of the constant scalar curvature slices in Schwarzschild solutions. The left panel depicts the positive mass case. The right panel corresponds to the negative mass Schwarzschild solution.}
\end{figure}

Each Schwarzschild solution, with both positive and negative Schwarzschild mass $M$, contains a large family of spherically symmetric CMC 3-slices. These are discussed at length for example in \cite{MOM}. A special member of these CMC slices is the so-called `umbilical' slice, where the extrinsic curvature $K^{ij}$ is proportional to the metric $g_{ij}$. In the CMC case this means that $K_{ij} = \frac{1}{3}Kg_{ij}$, where $K$ is the mean curvature (the trace of the extrinsic curvature), and is a constant. Using the Schwarzschild (areal) radius $R$ as a coordinate, we can explicitly write down the 3-metrics of each of these umbilical CMC slices as (see \cite{MOM})
\begin{equation}
\label{CMC}
dS^2 =  \frac{dR^2}{1 - \frac{2M}{R}  + \frac{K^2R^2}{9} } + R^2 d \Omega^2.
\end{equation}
Each of these metrics has constant negative scalar curvature $^{(3)}R$. We can see this immediately from the Hamiltonian constraint $^{(3)}R = K^{ij}K_{ij} - K^2 = -2 K^2 / 3$. The embedding pictures in Penrose diagrams are shown in Fig.~\ref{fig13}, for positive and negative masses $M$. In the positive mass case, with $K > 0$, the slice runs from future null infinity to future null infinity. Along each slice there is a minimal area two surface, a `throat', with Schwarzschild radius $R_t < 2M$. The denominator in the metric given by Eq.~(\ref{CMC}) can be expressed as
\begin{equation}
\label{Rt}
  1 - \frac{2M}{R}  + \frac{K^2R^2}{9}  = \left[\frac{dR}{dS}\right]^2,
\end{equation}
where $dS$ is the proper distance along the slice. The `throat' is defined by $dR/dS = 0$. Therefore
\begin{equation}
\label{Rt1}
  1 - \frac{2M}{R_t}  + \frac{K^2R_t^2}{9}  = 0,
\end{equation}
or
\begin{equation}
\label{Rt2}
   \frac{K^2R_t^3}{9}  + R_t - 2M = 0.
\end{equation}
It is straightforward to show that the cubic equation, Eq.~(\ref{Rt2}), has a single positive root $R_t$ satisfying $R_t < 2M$.

When the Schwarzschild mass $M$ is negative, the umbilical CMC slice runs from future null infinity, where $R \rightarrow \infty$, directly into the naked singularity at $R = 0$ and $R$ decreases monotonically along the slice.

It turns out that the 2-parameter family of metrics given by Eq.~(\ref{CMC}), where the parameters are $M$ and $K^2$, represent \textit{all} spherical constant negative scalar curvature 3-manifolds. We know that every spherical 3-metric can be written in conformally flat coordinates. The immediate challenge of this paper is to find conformal factors $\phi(r)$ so as to write each metric of Eq.~(\ref{CMC}) as
\begin{equation}
\label{CMC1}
dS^2 =  \frac{dR^2}{1 - \frac{2M}{R}  + \frac{K^2R^2}{9} } + R^2 d \Omega^2 = \phi^4\left(dr^2 + r^2d \Omega^2\right) .
\end{equation}

\section{Constant negative scalar curvature spherical 3-metrics in conformal coordinates}

It turns out that if we are given a conformally flat metric $g_{ij} = \phi^4 \delta_{ij}$, as in Eq.~(\ref{CMC1}) (here $\delta_{ij}$ denotes the flat metric tensor), the scalar curvature of the conformally flat metric ${^{(3)}R}$ satisfies
\begin{equation}
\label{cf}
8\nabla^2 \phi +  {^{(3)}R} \, \phi^5 = 0,
\end{equation}
where $\nabla^2$ is the flat space Laplacian. This is the key equation. We want to solve it assuming that ${^{(3)}R} = -\frac{2}{3}K^2$ is a negative constant. Therefore Eq.~(\ref{cf}) becomes
\begin{equation}
\label{cf1}
8\nabla^2 \phi - \frac{2}{3} K^2 \, \phi^5 = 0.
\end{equation}
This equation has an obvious rescaling property. If we have a solution $\phi$, and we multiply it by a constant $\lambda$, we get another solution with the coefficient $2 K^2/3$ divided by $\lambda^4$. This allows us to set the coefficient equal to any constant we want (8 is the obvious choice), and simplify the equation to
\begin{equation}
\label{cf2}
\nabla^2 \phi -  \phi^5 = 0
\end{equation}
and, once we have a solution, we can, by rescaling, find a solution for any constant negative value of the scalar curvature.

Aside: Such a rescaling is a legitimate transformation because, as we will see, all solutions of Eq.~(\ref{cf1}) or Eq.~(\ref{cf2}) either blow up or go to zero at the infinity so there is no natural `boundary condition' there which we should impose, and which would be disturbed by such a rescaling. This reduces the set of solutions from depending on two parameters $(M, K^2)$ to depending only on one, $M$. 

We seek spherically symmetric solutions. Therefore, in spherical polar coordinates we will write Eq.~(\ref{cf2}) as
\begin{equation}
\label{cf3}
\frac{d^2\phi}{dr^2} + \frac{2}{r}\frac{d\phi}{dr} -  \phi^5 = 0.
\end{equation}
This equation has another rescaling invariance. Let us have a solution $\phi(r)$. Pick any constant $\lambda$,  and rescale the coordinate $r$ and the function $\phi$ via 
\begin{equation}
\label{lambda}
r' = r/\lambda^2 \hskip 0.5cm{\rm and} \hskip 0.5cm \phi'(r') = \phi(r)/\lambda.
\end{equation}
Now the new function $\phi'$ also satisfies Eq.~(\ref{cf3}) relative to the new coordinate $r'$, with the {\it same} value of the scalar curvature. This means that we do not expect to have a unique solution to Eq.~(\ref{cf2}) or Eq.~(\ref{cf3}) corresponding to each value of $M$. Rather we will have an infinite family, generated by this rescaling.
 
We can immediately write down two solutions to Eq.~(\ref{cf3}). These are
 \begin{equation}
 \label{phi1}
 \phi_1 = \frac{\sqrt[4]{3}\sqrt{\alpha}}{\sqrt{\alpha^2 - r^2}}, \hskip  1cm \phi_2 = \frac{\sqrt[4]{3}\sqrt{\beta}}{\sqrt{ r^2 - \beta^2}}, 
\end{equation} 
where $\alpha$ and $\beta$ are arbitrary constants. Changing the value of $\alpha$ (or $\beta$) generates a rescaling transformation of the kind defined by Eq.~(\ref{lambda}).

The rescaling freedom of Eq.~(\ref{lambda}) suggests the following change of variables\footnote{This transformation was pointed out to us by Piotr Bizo\'{n}. It is the key to solving this problem explicitly.}
\begin{equation}
\label{zt}
z = \phi\sqrt{r} \quad \mathrm{and} \quad t = -\ln r.
\end{equation}
First, write $\phi(r) = z(r)/\sqrt{r}$. We get
\begin{eqnarray*}
\phi'&= &z'/\sqrt{r} - z/2r^{3/2},\\
\phi'' &=& z''/\sqrt{r} -z'/r^{3/2} + 3z/4r^{5/2},\\
2\phi'/r& = &2z'/r^{3/2} - z/r^{5/2}.\\
\end{eqnarray*}
We then have
\begin{equation}
\label{zt1}
\phi'' +2\phi'/r - \phi^5 = z''/\sqrt{r} +z'/r^{3/2} - z/4r^{5/2} - z^5/r^{5/2}.
\end{equation}
We now change the variable from $r$ to $t = -\ln r$. We choose this notation specially because we want to view the system as a particle moving in a potential. This yields
\begin{equation}
t = -\ln r \Rightarrow dt = -dr/r, \quad dr/dt = -r,
\end{equation}
and
\begin{equation}
dz/dt  = -rdz/dr, \quad d^2z/dt^2 = r^2d^2z/dr^2 +rdz/dr.
\end{equation}
Therefore Eq.~(\ref{zt1}) simplifies to 
\begin{equation}
\label{zt2}
d^2z/dt^2 - z/4 -z^5 = 0.
\end{equation}
This equation represents a particle moving along the $z$ axis under the influence of a repulsive potential $V = z/4 + z^5$. This form of the equation has a simple first integral
\begin{equation}
\label{M}
\left( \frac{dz}{dt} \right)^2 =  \frac{ z^2 }{4} +  \frac{z^6}{3}  - \frac{M}{2},
\end{equation}
where $-M/4$ can be regarded as the total energy of the `particle'.\footnote{Eq.~(\ref{cf3}) has a first integral $r^3\phi'^2 - r^3\phi^6/3 + r^2\phi\phi'$ (Rick Schoen, private communication). Expression Eq.~(\ref{M}) is this conserved quantity in the new variables.} Note that we can add an arbitrary constant to the definition of `time'. This is equivalent to multiplying $r$ by a constant and thus generates the rescaling of Eq.~(\ref{lambda}). We have chosen notation for the total energy deliberately. It will emerge that this $M$ coincides with the value of the mass of the Schwarzschild solution this given constant negative scalar curvature can be imbedded into.

We can, and will, write down the explicit solutions of Eq.~(\ref{M}), and in turn the solutions of Eq.~(\ref{cf3}), in terms of Weierstrass elliptic functions. However, before we do so we would like to study the qualitative nature of the solutions. The repulsive potential increases rapidly with increasing $z$. This means that the `particle' will reach infinity in finite `time', irrespective of the value of $M$. Hence the value of $\phi$ will always blow up at some finite $r$.

The solutions split into three groups, depending on the sign of $M$. There are the solutions with $M > 0$. These come in from $z =\infty$ starting at some finite time, strike the potential barrier at some strictly positive value of $z$, and return to $z = \infty$. When we translate back to $\phi(r)$, we get a `U' shaped solution. It starts off with $\phi = \infty$ at some value of $r$, decreases as $r$ decreases until $\phi$ reaches some finite minimum value and increases again to reach infinity at some other smaller value of $r$. Note that we chose $t = -\ln r$, so $r$ decreases as $t$ increases. The minimum value of $z$, call it $z_m$, occurs when
 \begin{equation}
 \label{zm}
\frac{dz}{dt}  =  0 \Rightarrow \frac{ z_m^2 }{4} +  \frac{z_m^6}{3}  -\frac{M}{2} = 0.
\end{equation}
Remember that $z^2 = \phi^2r$, but of course $\phi^2r = R$, the Schwarzschild radius. Therefore the minimum value of $z = z_m$ coincides with the minimum value of $R = R_t$ such that $z^2_m = R_t$. Therefore Eq.~(\ref{zm}) reads $R_t^3/3 + R_t/4 - M/2 = 0$. The equation that $R_t$ satisfies, in the umbilical slice, is $K^2R_t^3/9  + R_t - 2M = 0$ (Eq.~(\ref{Rt2})). We have set ${^{(3)}}R = -2K^2/3 = -8$, so $K^2 = 12$. This equation becomes $4R_t^3/3 + R_t - 2M = 0$. By comparing the two equations  we immediately see that $-M/2$ is the correct expression for the potential. Therefore the `U' shaped solutions are the umbilical slices which run from one future infinity to the other. We see that as $z$ diverges, $R$ diverges, and $\phi$ diverges, but $t$ runs over a finite interval and $r$, the flat radial coordinate, only covers a similar finite interval.
  
The solutions with $M < 0$ are the positive energy solutions and correspond to the umbilical slices in the negative mass Schwarzschild solution. They do not `bounce' off the potential, but continue all the way into $z = 0$, which, of course corresponds to $R = 0$. There are two sets of solutions. One set comes in from infinity and strike the singularity at $R = 0$ while the other starts at $R = 0$ and heads out to infinity.
  
Finally, there are the special cases where $M = 0$. These generate slices which can be embedded in Minkowski space. One family corresponds to a particle which starts off with $z = \infty$ at some finite time but will take an infinite amount of time to reach $z = 0$. This corresponds to the solutions $\phi_1$ of Eq.~(\ref{phi1}). The other solutions are just the time-reversal. One starts off at $z = 0$ at $t = -\infty$ and takes an infinite time to get to $z = \infty$, but arrives there at a finite $t$, i.e., nonzero $r$. These are the solutions $\phi_2$ of Eq.~(\ref{phi1}).
  
To derive the solutions in closed form, we rewrite Eq.~(\ref{M}) as
\[  2 \sqrt{3} \frac{dz}{\sqrt{4 z^6 + 3 z^2 - 6M}} = \pm dt. \]
The polynomial
\[ w(z) = \frac{1}{3} z^6 + \frac{1}{4} z^2 - \frac{M}{2} \]
can be factorised. One gets
\[ w(z) = \frac{1}{3} (z^2 - a) \left( z^4 + a z^2 + a^2 + \frac{3}{4} \right),  \]
where
\[ a = \frac{1}{2} \left( A - \frac{1}{A} \right), \;\;\; A = \left( \sqrt{1 + 36M^2}  + 6M \right)^{1/3}.  \]
Thus, for $M >  0$, the polynomial $w(z)$ is positive for $z \in (- \infty, -\sqrt{a}) \cup (\sqrt{a}, \infty)$. For $M \le 0$, $w(z)$ is positive on the entire real axis. The integral
\[ I = 2 \sqrt{3} \int \frac{dz}{\sqrt{4 z^6 + 3 z^2 - 6M}} \]
can be converted from a sextic to a cubic integral by the substitution $z = \pm \sqrt{6M/(1 - 12y)}$. This yields
\begin{eqnarray*}
I & = & \mp \int \frac{dy}{\sqrt{4 y^3 - \frac{1}{12}y + \left( \frac{1}{3} M^2 + \frac{1}{216} \right)}} \\
& = & \wp^{-1} \left( y; \frac{1}{12}, - \left( \frac{1}{3} M^2 + \frac{1}{216} \right) \right) + C,
\end{eqnarray*}
where $\wp(y;g_2,g_3)$ is the Weierstrass elliptic function with weights $g_2$ and $g_3$, and $C$ is a constant. The corresponding solution for $\phi$ can be written as
\begin{equation} \phi = \pm \frac{1}{\sqrt{r}} \sqrt{\frac{M/2}{\frac{1}{12} - \wp \left( \ln(Br); \frac{1}{12}, - \left( \frac{1}{3}M^2 + \frac{1}{216} \right) \right) }}, \end{equation}
but, making use of Eq.~(\ref{appendix1}), one can also write
\begin{equation}
\label{aabb}
\phi = \pm \frac{1}{\sqrt{r}} \sqrt{\frac{6M}{1 - \wp \left( \frac{\ln(Br)}{2 \sqrt{3}}; 12, - 8 \left( 72M^2 + 1 \right) \right) }}.
\end{equation}
Here $B$ is an integration constant. The above solution is valid both for  positive and negative $M$. For positive $M$ the domain of $\phi$ is given by
\[ \wp \left( \frac{\ln(Br)}{2 \sqrt{3}}; 12, - 8 \left( 72M^2 + 1 \right) \right) < 1.  \]
For negative $M$ the above inequality is reversed.  For $M = 0$ we have two distinct families of solutions, given by Eq.~(\ref{phi1}). They can also be written as
\[ \phi = \pm \sqrt{\frac{B}{B^2 - \frac{1}{3}r^2}}, \quad \phi = \pm \sqrt{\frac{D}{\frac{1}{3}r^2 - D^2}}, \]
where $B$ and $D$ are integration constants.

\begin{figure}
\includegraphics[width=11cm]{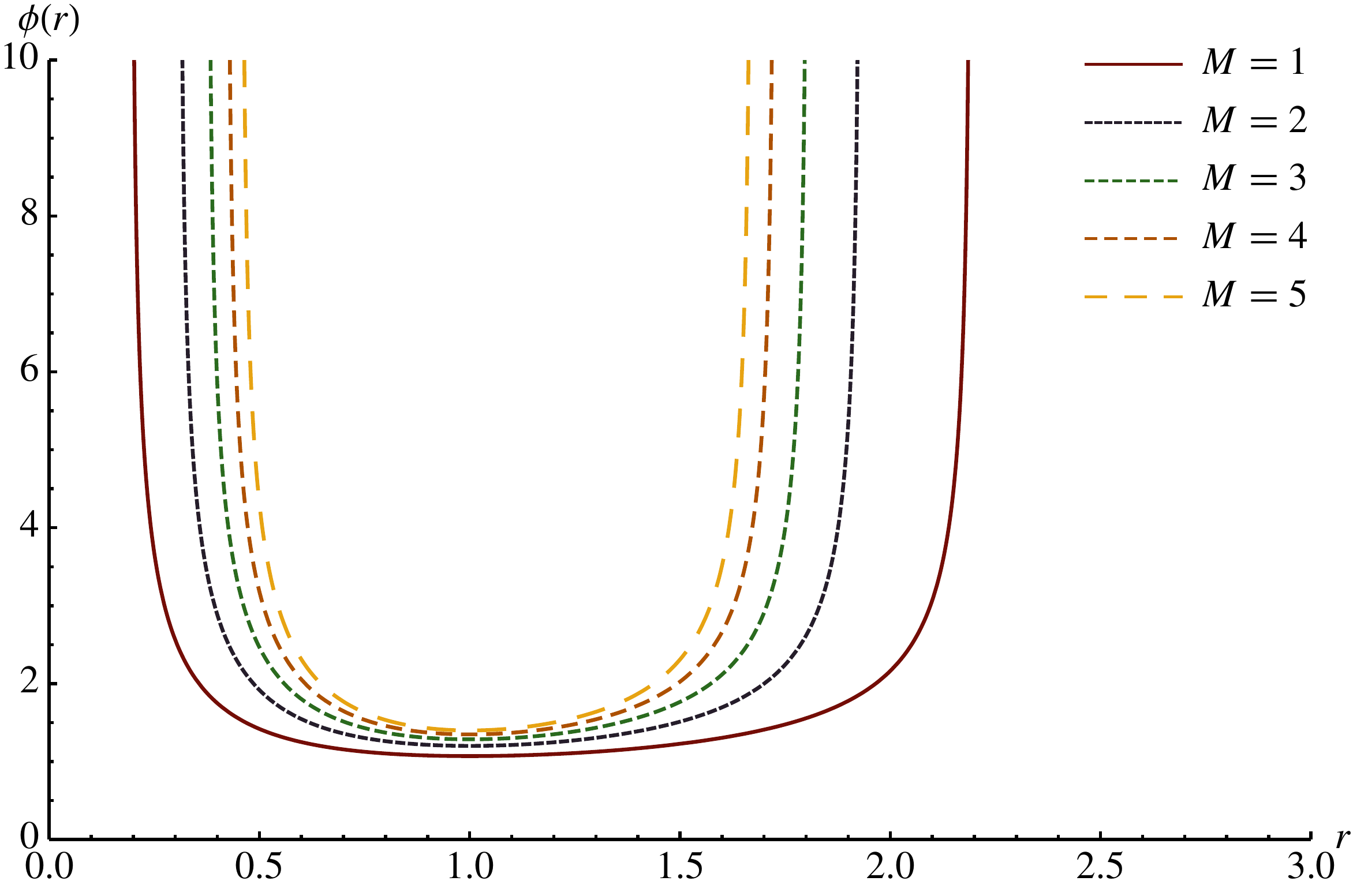}
\caption{\label{fig1}Solutions (\ref{aabb}) with positive constants $M = 1,2,3,4,$ and 5. The values of $B$ are set to $B \approx 0.061869$, 0.102657, 0.133031, 0.157389, and 0.177778, respectively. This choice assures that each of the above solutions has a minimum at $r = 1$.}
\end{figure}

\begin{figure}
\includegraphics[width=11cm]{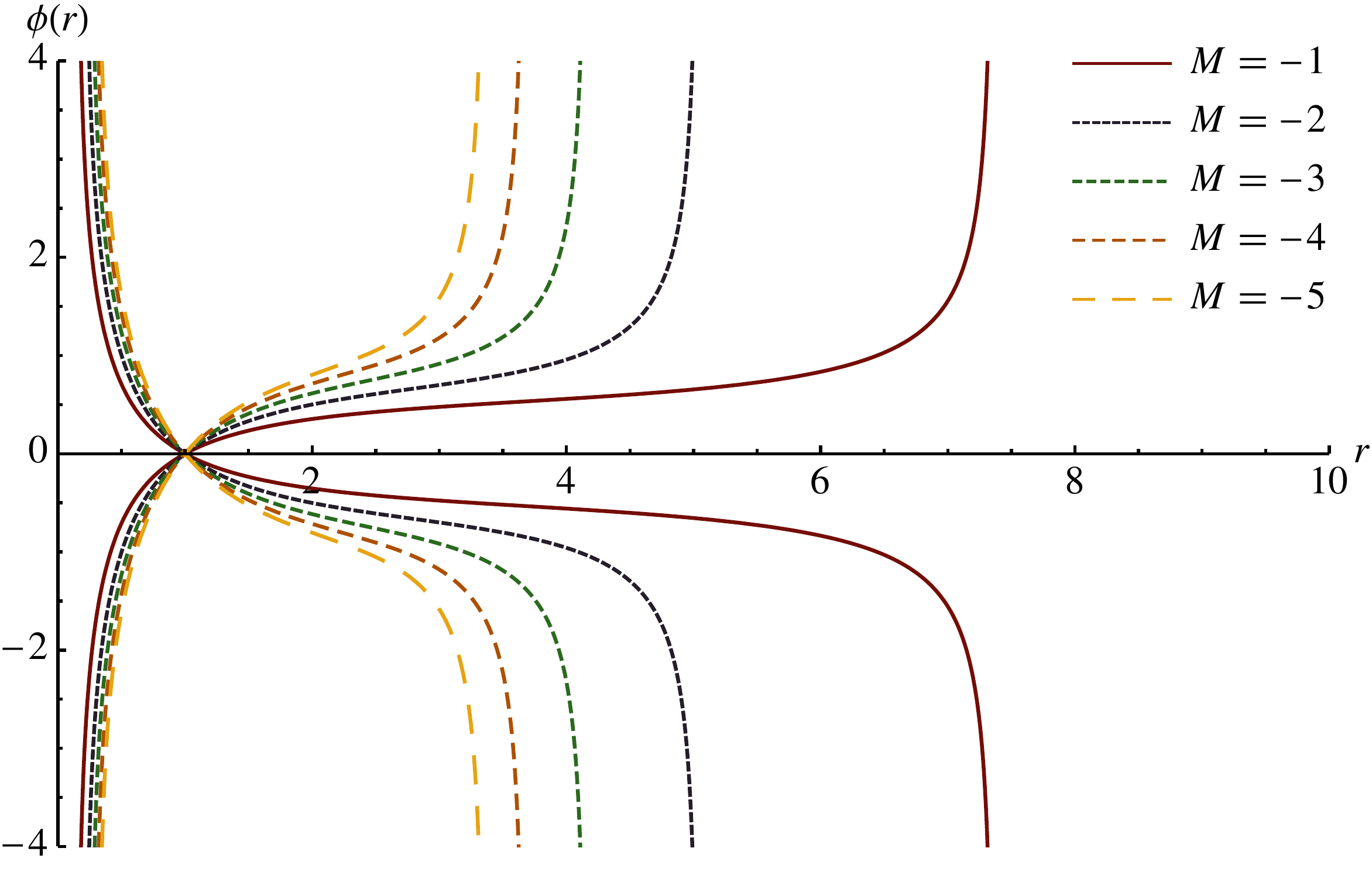}
\caption{\label{fig2}Solutions (\ref{aabb}) with negative constants $M = -1$, -2, -3, -4, and -5. Here $B = 1$.}
\end{figure}

Both the positive and negative mass solutions can be easily plotted using, say, \textit{Wolfram Mathematica}. We construct representative families of solutions for both positive and negative $M$. They are plotted in Figs.~\ref{fig1} and \ref{fig2}, respectively.

The equation corresponding to positive scalar curvature is
\[ \frac{d^2\phi}{dr^2} + \frac{2}{r}\frac{d\phi}{dr} +  \phi^5 = 0, \]
which is one of Lane--Emden equations. It appears frequently in the astrophysical literature, due to its application in the modelling of static configurations of self-gravitating perfect fluids (stars) or globular clusters. All solutions of this equation are listed in \cite{mach_jmp}. Some $N$-dimensional generalisations of Lane--Emden equations were also considered in the literature, e.g., in \cite{horedt, horedt_book}. Thus, in particular, Eq.~(\ref{eq3}) of this paper corresponds to Eq.~(65) of \cite{horedt}.

\section{Constant scalar curvature metrics in polar coordinates}

Spherically symmetric Riemannian metrics of constant scalar curvature can be obtained for any dimension $N \ge 2$ in polar coordinates. 

Let $N = n+1 \ge 3$ (the 2-dimensional case can be considered separately, but it is not of much interest to our discussion). The line element on an $N$-dimensional spherically symmetric manifold can be written as
\begin{equation}
\label{spherical}
ds^2 = d\zeta^2 + f^2(\zeta) d \Omega_n^2,
\end{equation}
where $d\Omega_n^2$ denotes the line element of a round metric on an $n$-dimensional sphere. A straightforward (but lengthy) calculation  shows that the Ricci scalar corresponding to the line element (\ref{spherical}) has the form
\[ {^{(N)}R} = \frac{1}{f^2} \left( -2 n f f^{\prime \prime} - n (n-1) {f^{\prime}}^2 + R_n \right), \]
where $R_n = n(n-1)$ is the scalar curvature of an $n$-sphere. For constant ${^{(N)}R}$, the equation
\[ {^{(N)}R} f^2 = -2 n f f^{\prime \prime} - n(n-1) {f^\prime}^2 + n(n-1)  \]
can be integrated at least once. Multiplying the above equation by $f^{n-2} f^\prime$ and rearranging terms, we get
\[ \left( (n-1){f^\prime}^2 + 2 f f^{\prime \prime}  \right) f^{n-2} f^\prime =    \left( f^{n-1} {f^\prime}^2 \right)^\prime = \left( (n-1) f^{n-2} - \frac{{^{(N)}R}}{n} f^n \right) f^\prime,  \]
and thus
\[ {f^\prime}^2 = 1 - \frac{{^{(N)}R}}{n(n+1)} f^2 + C_1 f^{1 - n},  \]
where $C_1$ denotes an integration constant. Surprisingly, there is no need of searching for an explicit solution for $f$. Defining $\xi = f(\zeta)$, one obtains
\begin{eqnarray}
\label{ds}
ds^2 & = &  \frac{d\xi^2}{1 - \frac{{^{(N)}R}}{n(n+1)} \xi^2 + C_1 \xi^{1-n} } + \xi^2 d \Omega_n^2 \\
& = & \frac{d\xi^2}{1 - \frac{{^{(N)}R}}{N(N-1)} \xi^2 + C_1 \xi^{2-N} } + \xi^2 d \Omega_{N-1}^2. \nonumber
\end{eqnarray}
Note that for $C_1 = 0$ and ${^{(N)}R} = N(N-1)$ the above expression yields a round metric on an $N$-sphere. To see this, it is enough to substitute $\xi = \sin r$. In this case
\[ ds^2 = dr^2 + \sin^2 (r) d\Omega^2_{N-1} = d\Omega_N^2. \]
Similarily, for $C_1 = 0$, and ${^{(N)}R} = -N(N-1)$, substitution $\xi = \sinh r$ yields a metric of a hyperboloid
\[  ds^2 = dr^2 + \sinh^2 (r) d\Omega^2_{N-1}. \]

\section{Conformally flat form in higher dimensions}

The convenient form of the conformal transformation in $N \ge 3$ dimensions is to set $g_{ij} = \phi^{4/(N-2)} \delta_{ij}$. This yields
\begin{equation}
\label{aaa}
\frac{4 (N-1)}{N-2} \nabla^2 \phi + {^{(N)}R} \phi^{(N+2)/(N-2)} = 0,
\end{equation}
which is a straightforward generalisation of Eq.~(\ref{cf}). Here ${^{(N)}R}$ is the Ricci scalar corresponding to $g_{ij}$, and $\nabla^2$ denotes the flat Laplacian. The above equation admits a scaling symmetry. If $\phi$ is a solution corresponding to the scalar curvature ${^{(N)}R}$, then $\lambda \phi$ is a solution corresponding to the curvature ${^{(N)}R}/\lambda^{4/(N-2)}$, for any $\lambda \in \mathbb R_{+}$.

As before, for a constant scalar curvature ${^{(N)}R}$, Eq.~(\ref{aaa}) can be simplified to
\begin{equation}
\label{eq1}
\nabla^2 \phi + \epsilon \phi^{(N+2)/(N-2)} = 0,
\end{equation}
where $\epsilon = \mathrm{sign} \left({^{(N)}R} \right)$.

In spherical symmetry, Eq.~(\ref{eq1}) can be written as
\begin{equation}
\label{aa}
\frac{d^2 \phi}{dr^2} + \frac{N-1}{r} \frac{d\phi}{dr} + \epsilon \phi^{(N+2)/(N-2)} = 0,
\end{equation}
where $r$ denotes the radial coordinate. Equations (\ref{eq1}) and (\ref{aa}) are invariant under the scaling transformation
\begin{equation}
\label{scaling}
\phi(x^i) \mapsto \phi(x^i/\lambda)/\lambda^{(N-2)/2},
\end{equation}
with $\lambda \in \mathbb R_{+}$. This suggests the substitution $\phi = r^{-(N-2)/2} z$, $t = -\ln r$ in Eq.~(\ref{aa}). It yields
\begin{equation}
\label{dd}
\frac{d^2 z}{dt^2} = \left( \frac{N-2}{2} \right)^2 z - \epsilon z^{(N+2)/(N-2)}.
\end{equation}
A standard integration of the above equation gives
\begin{equation}
\label{bb}
\left( \frac{dz}{dt} \right)^2 = \left( \frac{N-2}{2} \right)^2 z^2 - \epsilon \frac{N-2}{N} z^{2N/(N-2)} + C,
\end{equation}
or
\begin{equation}
\label{eq2}
\frac{dz}{\sqrt{ \left( \frac{N-2}{2} \right)^2 z^2 - \epsilon \frac{N-2}{N} z^{2N/(N-2)} + C  }} = \pm dt.
\end{equation}

Equation (\ref{eq2}) can be integrated for all $N \ge 3$ in a special case of $C = 0$. The only solutions that are regular at $r = 0$ belong to that class. For arbitrary values of $C$ one can find analytic solutions if
\[ w(z) \equiv \left( \frac{N-2}{2} \right)^2 z^2 - \epsilon \frac{N-2}{N} z^{2N/(N-2)} + C \]
appearing in Eq.~(\ref{eq2}) is a polynomial. This happens for $N = 3, 4$ and 6 only (note that these are the only integers $N \ge 3$ for which the expression $(N+2)/(N-2)$ is also an integer). Solutions obtained in those cases are mostly given in terms of elliptic functions, but there are exceptions --- for some special values of $C$ solutions can be expressed in terms of elementary functions.

The integration constant $C$ can be redefined so that it corresponds to the mass of the suitable $N$-dimensional Schwarzschild–-Tangherlini solution (see, e.g., \cite{emparan_reall}). We are convinced, however, that such a redefinition would add unnecessary complication to this paper.

\subsection{Relation with the polar metric}

Equation (\ref{eq2}) can be also obtained directly from Eq.~(\ref{ds}).  The line element (\ref{ds}) can be written in a conformally flat form by a suitable change of coordinates $\xi = g(r)$, $d\xi = g^\prime(r) dr$, i.e.,
\begin{eqnarray*}
ds^2 & = &  \frac{d\xi^2}{1 - \frac{{^{(N)}R}}{N(N-1)} \xi^2 + C_1 \xi^{2-N} } + \xi^2 d \Omega_{N-1}^2 \\
& = & \frac{{g^\prime}^2 dr^2}{1 - \frac{{^{(N)}R}}{N(N-1)} g^2 + C_1 g^{2-N} } + g^2 d \Omega_{N-1}^2 = \phi^\frac{4}{N-2}(r) \left( dr^2 + r^2 d \Omega_{N-1}^2 \right).
\end{eqnarray*}
The resulting conditions
\[ \frac{{g^\prime}^2}{1 - \frac{{^{(N)}R}}{N(N-1)} g^2 + C_1 g^{2-N}} = \phi^\frac{4}{N-2}, \;\;\; g^2 = \phi^\frac{4}{N-2} r^2 \]
yield
\[ \frac{dg}{\sqrt{g^2 - \frac{{^{(N)}R}}{N(N-1)} g^4 + C_1 g^{4-N} }} = \pm dt, \]
where $t = - \ln r$, and $g = r \phi^{2/(N-2)} = z^{2/(N-2)}$, according to the notation used in the introduction of this paper. The latter relation allows us to write
\[ \frac{dz}{\sqrt{\left( \frac{N-2}{2} \right)^2 z^2 - \frac{{^{(N)}R}}{N(N-1)} \left( \frac{N-2}{2} \right)^2 z^\frac{2N}{N-2}  + \left( \frac{N-2}{2} \right)^2 C_1}} = \pm dt. \]
The above equation is identical with Eq.~(\ref{eq2}) where $C = ((N-2)/2)^2 C_1$ and $\epsilon$ replaced by ${^{(N)}R}(N-2)/(4(N-1))$ (cf.~Eq.~(\ref{aaa})).

In the following, symbols $A$ and $B$ will be reserved for integration constants.

\section{Solutions with $C=0$}

The integral
\begin{eqnarray*}
I & = & \int \frac{dz}{\sqrt{ \left( \frac{N-2}{2} \right)^2 z^2 - \epsilon \frac{N-2}{N} z^{2N/(N-2)} + C  }} \\
& = & \int \frac{dz}{|z| \sqrt{\left(\frac{N-2}{2} \right)^2 - \epsilon \frac{N-2}{N} z^{4/(N-2)}}}
\end{eqnarray*}
can be computed by substituting
\[ x = \frac{2}{N-2} \sqrt{\left( \frac{N-2}{2}\right)^2 - \epsilon \frac{N-2}{N} z^{4/(N-2)}}. \]
This yields
\[ I = \pm \int \frac{dx}{1 - x^2} = \pm \left\{ \begin{array}{l} \mathrm{ar \, tanh} x \\ \mathrm{ar \, ctanh} x \end{array}\right. + A. \]
The corresponding solution of Eq.~(\ref{aa}) has the form
\begin{equation}
\label{eq3}
\phi = \alpha \left( \delta \frac{r^2}{N(N-2)} + \beta \right)^{(2-N)/2},
\end{equation}
where $\alpha = \pm 1$, $\beta = \pm 1$, $\delta = \pm 1$, and $\epsilon = \alpha^{4/(N-2)} \beta \delta$. Solutions (\ref{eq3}) are given in a normalised form --- those that are regular at $r = 0$ satisfy $\phi(r=0) = \pm 1$. The whole one parameter family of solutions can be obtained from each solution (\ref{eq3}) by means of the scaling transformation (\ref{scaling}).

\section{Power-law solutions}
\label{sec_power_law}

It is clear from Eq.~(\ref{dd}), that it admits solutions of the form $z \equiv z_0$, where $z_0$ is a root of
\[ \left( \frac{N-2}{2} \right)^2 - \epsilon z^{4/(N-2)} = 0. \]
For instance, $z \equiv (N-2)/2)^{(N-2)/2}$ is always a solution of Eq.~(\ref{dd}) with $\epsilon = +1$, and the corresponding expression for $\phi$ is
\[ \phi = \left( \frac{(N-2)/2}{r} \right)^{(N-2)/2}. \]

\section{Solutions for $N=4$}

For $N = 4$ Eq.~(\ref{aa}) has the form
\begin{equation}
\label{cc}
\frac{d^2 \phi}{dr^2} + \frac{3}{r} \frac{d \phi}{dr} + \epsilon \phi^3 = 0.
\end{equation}

\subsection{Solutions for $\epsilon = +1$}

With $N = 4$ and $\epsilon = +1$, Eq.~(\ref{eq2}) can be written as
\begin{equation}
\label{eq4}
\frac{dz}{\sqrt{- \frac{1}{2} z^4 + z^2 + C}} = \pm dt.
\end{equation}

\tolerance=1000
The classification of solutions follows from the properties of
\[ w(z) = -\frac{1}{2}z^4 + z^2 + C = \frac{1}{2} z^2 \left(\sqrt{2} - z \right) \left(\sqrt{2} + z\right) + C.\]
\begin{enumerate}
\item For $C < 1/2$ the polynomial $w(z)$ is always negative --- there are no real solutions of Eq.~(\ref{eq4})
\item For $C = -1/2$ the polynomial $w(z)$ has two real zeros at $z = \pm 1$, and it is non-positive. Functions $z \equiv \pm 1$ solve Eq.~(\ref{bb}). They correspond to solutions
\[ \phi = \pm 1/r \]
\item For $C \in (-1/2,0)$ the polynomial $w(z)$ has four real roots: $\pm \sqrt{1 - \sqrt{1 + 2C}}$ and $\pm \sqrt{1 +\sqrt{1 + 2C}}$. It is positive for $\sqrt{1 - \sqrt{1 + 2C}} < |z| < \sqrt{1 + \sqrt{1 + 2C}}$
\item For $C=0$ the polynomial $w(z)$ has three real zeros: at $z = \pm \sqrt{2}$ and $z = 0$. This case yields a regular solution of the form given by Eq.~(\ref{eq3}), that is
\[ \phi = \pm \frac{B}{B^2 + \frac{1}{8} r^2} \]
\item For $C > 0$ the polynomial $w(z)$ has two real zeros at $\pm \sqrt{1 + \sqrt{1 + 2C}}$. It is positive for $|z| < \sqrt{1 + \sqrt{1 + 2C}}$
\end{enumerate}

\subsubsection{$C \in (-1/2,0)$}

For $C \in (-1/2,0)$ we have $w(z) = \frac{1}{2}(z^2 - a)(b - z^2)$, where $a = 1 - \sqrt{1 + 2C}$, $b = 1 + \sqrt{1 + 2C}$. Here $0 < a < b$. Equation (\ref{eq4}) can be written as
\[ \frac{\sqrt{2} dz}{\sqrt{z^2 - a} \sqrt{b - z^2}} = \pm dt, \]
where $a < z^2 < b$. It can be integrated by substituting $y = z/\sqrt{b}$, $a/b < y^2 < 1$. This yields
\begin{eqnarray*}
I & = & \int \frac{\sqrt{2} dz}{\sqrt{z^2 - a} \sqrt{b - z^2}} = \sqrt{\frac{2}{b}} \int \frac{dy}{\sqrt{1 - y^2} \sqrt{y^2 - \frac{a}{b}}} \\
& = & -\sqrt{\frac{2}{b}} \mathrm{arc \, dn} \left(y, \sqrt{\frac{b - a}{b}}\right) + A.
\end{eqnarray*}
Here $\mathrm{arc \, dn}$ is the inverse of the Jacobian elliptic function $\mathrm{dn}$. The integral representations of all inverse Jacobian elliptic functions appearing in this paper are collected in the Appendix. Returning to the original variables, one obtains a solution of Eq.~(\ref{cc}) in the form
\[ \phi = \pm \frac{\sqrt{1 + K}}{r} \mathrm{dn} \left( \sqrt{ \frac{1 + K}{2}} \mathrm{ln}(Br), \sqrt{\frac{2K}{1 + K}} \right), \;\;\; K = \sqrt{1 + 2C}. \]

\subsubsection{$C > 0$}

For $C > 0$ we can factorise $w(z)$ as $w(z) = \frac{1}{2}(a - z^2)(z^2 + b)$, where $a = 1 + \sqrt{1 + 2C}$, $b = \sqrt{1 + 2C} - 1$. Here $0 < b < a$, and $w(z)$ is positive for $z^2 < a$. Equation (\ref{eq4}) can be written in the form
\[ \frac{\sqrt{2} dz}{\sqrt{a-z^2}\sqrt{z^2 + b}} = \pm dt, \]
that can be integrated by substituting $y = z/\sqrt{a}$, $y^2 < 1$. We have
\begin{eqnarray*}
I & = & \int \frac{\sqrt{2} dz}{\sqrt{a-z^2}\sqrt{z^2 + b}} = \sqrt{2} \int \frac{dy}{\sqrt{1-y^2} \sqrt{b + ay^2}}\\
& = & - \sqrt{\frac{2}{a+b}} \mathrm{arc \, cn}\left(y, \sqrt{\frac{a}{a+b}} \right) + A.
\end{eqnarray*}
Here, similarly, $\mathrm{arc \, cn}$ denotes the inverse of the Jacobian function $\mathrm{cn}$. The solution can be now written as
\[ \phi = \pm \frac{\sqrt{1 + K}}{r} \mathrm{cn} \left( \sqrt{K} \mathrm{ln}(Br), \sqrt{\frac{1 + K}{2K}} \right), \;\;\; K = \sqrt{1 + 2C}. \]

\subsection{Solutions with $\epsilon = -1$}   

In this case real solutions exist for any value of $C$. The quartic polynomial $w(z) = \frac{1}{2} z^4 + z^2 + C$ is positive for $C >0$, and has two real zeros for $C < 0$. Equation (\ref{eq4}) can be written as
\[ \frac{dz}{\sqrt{\frac{1}{2} z^4 + z^2 + C}} = \pm dt. \]
An (almost) natural substitution $y = z^2 + 2/3$ yields
\begin{eqnarray*}
I & = & \int \frac{dz}{\sqrt{\frac{1}{2} z^4 + z^2 + C}} = \int \frac{\sqrt{2} dy}{\sqrt{4 y^3 - 8 \left(\frac{2}{3} - C \right) y - \frac{16}{3} \left(C - \frac{4}{9} \right) }} \\
& = & \sqrt{2} \wp^{-1} \left(y; 8 \left(\frac{2}{3} - C \right), \frac{16}{3} \left( C - \frac{4}{9} \right) \right) + A.
\end{eqnarray*}
The above result allows us to write the solution as
\[ \phi = \pm \frac{1}{r}\sqrt{\wp \left(\frac{\ln(Br)}{\sqrt{2}}; 8 \left( \frac{2}{3} - C \right), \frac{16}{3} \left(C - \frac{4}{9} \right) \right) - \frac{2}{3}} .\]

For many values of $C$ the solution can be obtained in terms of Jacobian elliptic functions or even as an elementary expression.

\subsubsection{$C < 0$}

In this case one has $w(z) = \frac{1}{2} z^4 + z^2 + C = \frac{1}{2}(z^2 - a)(z^2 + b)$, where $a = -1 + \sqrt{1 - 2C}$, $b = 1 + \sqrt{1 - 2C}$. Here $0 < a < b$. The polynomial $w(z)$ is clearly positive for $z^2 > a$. It follows that
\begin{eqnarray*}
I & = & \int \frac{dz}{\sqrt{\frac{1}{2} z^4 + z^2 + C}} = \int \frac{\sqrt{2} dz}{\sqrt{(z^2 - a)(z^2 + b)}}  \\
& = & \sqrt{\frac{2}{a+b}}\int \frac{dy}{\sqrt{(y^2 - 1) \left( \frac{a}{a+b} y^2 + \frac{b}{a+b} \right)}}\\
& = & \sqrt{\frac{2}{a+b}} \mathrm{arc \, nc} \left( y, \sqrt{\frac{b}{a+b}} \right) + A,
\end{eqnarray*}
where $y = z/\sqrt{a}$, $y^2 > 1$, and $\mathrm{arc \, nc}$ denotes the inverse of the elliptic function $\mathrm{nc}$.  The solution can be now written as
\[ \phi = \pm \frac{\sqrt{K-1}}{r} \mathrm{nc} \left( \sqrt{K} \ln (Br), \sqrt{\frac{1 + K}{2K}} \right), \;\;\; K = \sqrt{1 -2C}. \]

\subsubsection{$C=0$}

The solution is given by Eq.~(\ref{eq3}). In this particular case we get
\[ \phi = \pm \frac{B}{\frac{1}{8} r^2 - B^2}. \]

\subsubsection{$C \in (0,1/2)$}

Here
\begin{eqnarray*}
I & = & \int \frac{dz}{\sqrt{\frac{1}{2} z^4 + z^2 + C}} = \int \frac{\sqrt{2} dz}{\sqrt{(z^2 + a)(z^2 + b)}} \\
& = & \int \frac{\sqrt{2} dy}{\sqrt{b} \sqrt{(1 + y^2) \left( 1 + \frac{a}{b} y^2 \right)}} = \sqrt{\frac{2}{b}} \mathrm{arc \, sc} \left( y, \sqrt{\frac{b-a}{b}} \right) + A,
\end{eqnarray*}
where $a = 1 - \sqrt{1- 2C}$, $b = 1 + \sqrt{1 - 2C}$, $0 < a < b$, $y = z/\sqrt{a}$, $y \in \mathbb R$, and $\mathrm{arc \, sc}$ is the inverse of the elliptic function $\mathrm{sc}$. This form yields the following solution
\[\phi = \pm \frac{\sqrt{1 - K}}{r} \mathrm{sc} \left( \sqrt{\frac{1 + K}{2}} \ln(Br), \sqrt{\frac{2K}{1+K}} \right), \;\;\; K = \sqrt{1 - 2C}. \]

\subsubsection{$C = 1/2$}

In this case $w(z) = \frac{1}{2}(z^2 + 1)^2$. Thus
\[ I = \int \frac{dz}{\sqrt{\frac{1}{2} z^4 + z^2 + C}} = \int \frac{\sqrt{2} dz}{1 + z^2} = \sqrt{2} \mathrm{arc \, tan} \, z + A.  \]
The corresponding solution is
\[ \phi = \pm \frac{1}{r} \tan \left( \frac{\ln(Br)}{\sqrt{2}} \right). \]

\section{Solutions for $N=6$}

For $N = 6$, Eq.~(\ref{aa}) has the form
\[ \frac{d^2 \phi}{dr^2} + \frac{5}{r} \frac{d \phi}{dr} + \epsilon \phi^2 = 0. \]
It is enough to consider solutions with $\epsilon = +1$ or $\epsilon = -1$ only. If $\phi_1$ satisfies $\Delta \phi_1 - \phi_1^2 = 0$, then $\phi_2 = - \phi_1$ must be a solution to $\Delta \phi_2 + \phi_2^2 = 0$. We will derive solutions for $\epsilon = -1$, for convenience.

\subsection{Solutions for $\epsilon = - 1$}

In this case Eq.~(\ref{eq2}) can be written as
\begin{equation}
\label{eq5}
\frac{dz}{\sqrt{\frac{2}{3} z^3 + 4 z^2 + C}} = \pm dt,
\end{equation}
and it is clearly integrable in terms of the Weierstrass elliptic function $\wp$. Setting $y = z + 2$ we get
\begin{eqnarray*}
I & = & \int \frac{dz}{\sqrt{\frac{2}{3} z^3 + 4 z^2 + C}} = \sqrt{6} \int \frac{dy}{\sqrt{4y^3 - 48 y + 64 + 6 C}}\\
& = & \sqrt{3} \wp^{-1} \left( y; 48, -(64 + 6C) \right) + A.
\end{eqnarray*}
The corresponding solution has the form
\[ \phi = \frac{1}{r^2}\left( \wp \left( \frac{\ln (Br)}{\sqrt{6}}; 48, -(64 + 6 C) \right) - 2 \right). \]

A deeper insight into the properties of solutions of Eq.~(\ref{eq5}) is provided by the analysis of the polynomial $w(z) = \frac{2}{3} z^3 + 4 z^2 + C$.
\begin{enumerate}
\item There is one real, positive root of $w(z)$ for $C < -64/3$
\item For $C = -64/3$ the polynomial $w(z)$ has two zeros. One at $z = -4$, the other at $z = 2$. In this case $w(z)$ can be expressed as $w(z) = \frac{2}{3}(z+4)^2(z-2)$
\item There are three real roots of $w(z)$ for $C \in (-64/3, 0)$
\item For $C = 0$ the factorisation of $w(z)$ yields $w(z) = \frac{2}{3} z^2 (z + 6)$
\item For $C > 0$ the polynomial $w(z)$ has one real, negative root
\end{enumerate}

\subsubsection{$C = -64/3$}

Solutions for $C = -64/3$ can be given in terms of elementary functions. The first solution corresponds to a zero of $w(z)$ at $z = -4$. It is simply $\phi = - 4/r^2$, which is a power-law form discussed in Sec.~\ref{sec_power_law}. The other solution follows from the integration of
\[ \frac{\sqrt{3} dz}{\sqrt{2} (z + 4) \sqrt{z - 2}} = \pm dt. \]
Substituting $y = \sqrt{(z - 2)/6}$, one obtains
\[ I = \int \frac{\sqrt{3} dz}{\sqrt{2} (z + 4) \sqrt{z - 2}} = \int \frac{dy}{1 + y^2} = \mathrm{arc \, tan} \, y + A. \]
The corresponding solution is
\[ \phi = \frac{2}{r^2} \left( 1 + 3 \tan^2 ( \ln (Br)) \right). \]

\subsubsection{$C \in (-64/3,0)$}

Solutions for $C \in (-64/3,0)$ can be expressed by means of Jacobi elliptic functions. Define $K = \left( \frac{1}{4} \left( -32 - 3 C + \sqrt{3 C (64 + 3 C)} \right) \right)^{1/3}$, $a = K + 4/K$, $b = \frac{1}{2} \left( 3 a - \sqrt{3 (16 - a^2)} \right)$, and $c = \frac{1}{2} \left( 3 a + \sqrt{3 (16 - a^2)} \right)$. By substituting $z = b y^2 + a - 2$, one can show that
\begin{eqnarray*}
I & = & \int\frac{dz}{\sqrt{\frac{2}{3} z^3 + 4 z^2 + C}} = \frac{\sqrt{6}}{\sqrt{c}} \int \frac{dy}{\sqrt{1 + y^2} \sqrt{1 + \frac{b}{c} y^2}} \\
& = & \frac{\sqrt{6}}{\sqrt{c}}\mathrm{arc \, sc} \left( y, \sqrt{1 - \frac{b}{c}} \right) + A.
\end{eqnarray*}
The solution for the conformal factor has the form
\[ \phi = \frac{1}{r^2} \left( b \, \mathrm{sc}^2 \left( \frac{\sqrt{c}}{\sqrt{6}} \ln (B r) , \sqrt{1 - \frac{b}{c}} \right) + a - 2 \right). \]

\subsubsection{$C = 0$}

Solutions are given by Eq.~(\ref{eq3}), that is
\[ \phi = - \left( \frac{B}{B^2 + \frac{1}{24}r^2} \right)^2, \;\;\; \phi = \left( \frac{B}{B^2 - \frac{1}{24} r^2} \right)^2. \] 

\section{Generalised Srivastava functions}

It is quite surprising that in all dimensions considered so far, there is always a special solution that can be written down in terms of elementary functions. For $N = 3$ dimensions such a solution was first discovered by Srivastava \cite{srivastava}. It occurs for the Lane--Emden equation, that is for the positive scalar curvature, and can be written as
\[ \phi = \pm \frac{1}{\sqrt{r}} \left( 1 + 3 \tan^{-2} \left( \frac{1}{2} \ln (Br) \right) \right)^{-1/2} \]
(see also \cite{mach_jmp} for a discussion of how it fits into the formalism of this paper). The following solutions can be treated as generalisations of the original Srivastava's integral. For $N = 4$ and $\epsilon = -1$ we have
\[ \phi = \pm \frac{1}{r} \tan \left( \frac{1}{\sqrt{2}} \ln (Br) \right). \]
For $N = 6$ there is a solution of the form
\[ \phi = - 2 \epsilon \frac{1}{r^2} \left( 1 + 3 \tan^2 \left( \ln(Br) \right) \right). \]

\section{Summary}

We have obtained all spherically symmetric Riemannian metrics of constant scalar curvature in dimensions $N = 3, 4$ and 6 in the explicitly conformally flat form.

A few of presented solutions were known previously. Conformal factors obtained for positive scalar curvatures correspond to the solutions of the Lane--Emden equations with critical exponents. In $N = 3$ dimensions all such solutions were listed in \cite{mach_jmp}.

The resulting formulas use, in most cases, Jacobian or Weierstrass elliptic functions, but there are exceptions --- in some special cases such metrics can be written in terms of elementary functions as well. In dimensions $N=4$ and $N = 6$ we find special solutions that can be expressed in terms of simple trigonometric functions and that generalise the three dimensional solution discovered in \cite{srivastava}.

While the problem addressed in this paper can be thought of as being purely geometrical, three dimensional spherically symmetric Riemannian metrics of constant scalar curvature have a natural interpretation in General Relativity. They represent a special, simple choice of slices through the spacetime. An initial study of such slices through the Schwarzschild spacetime can be found in \cite{pareja}.

\section*{Acknowledgements}

One of us (PM) has been partially supported by the Polish Ministry of Science and Higher Education grant IP2012~000172 and the NCN grant DEC-2012/06/A/ST2/00397.

\section*{Appendix}

In the following we list all identities involving elliptic functions that were used in this paper. The adopted convention follows that of \cite{nist}. Symbols $k$ and $k^\prime$ are reserved for the modulus and the complementary modulus, respectively. They are related by $k^2 + {k^\prime}^2 = 1$. Moreover, $k \in [0,1]$, $k^\prime \in [0,1]$. We used the following integral representations of the inverse Jacobian elliptic functions:
\begin{eqnarray*}
\mathrm{arc \, dn}(x,k) & = & \int_x^1 \frac{dt}{\sqrt{\left(1-t^2\right) \left( t^2 - {k^\prime}^2 \right)}}, \;\;\; k^\prime \le x \le 1 \\
\mathrm{arc \, cn}(x,k) & = & \int_x^1 \frac{dt}{\sqrt{\left(1-t^2\right) \left( {k^\prime}^2 + k^2 t^2\right)}}, \;\;\; -1 \le x \le 1 \\
\mathrm{arc \, nc}(x,k) & = & \int_1^x \frac{dt}{\sqrt{\left(t^2 - 1\right)\left( k^2 + {k^\prime}^2 t^2 \right)}}, \;\;\; 1 \le x < \infty \\
\mathrm{arc \, sc}(x,k) & = & \int_0^x \frac{dt}{\sqrt{\left(1+t^2\right)\left( 1 + {k^\prime}^2 t^2 \right)}}, \;\;\; -\infty < x < \infty
\end{eqnarray*}

The Weierstrass elliptic function $\wp$ appears through the following relation:
\[ z = \int_{\wp(z;g_2,g_3)}^\infty \frac{dt}{\sqrt{4 t^3 - g_2 t - g_3}}. \]
Here the integral is taken along any path that does not pass through a zero of $4t^3 - g_2t - g_3$. The Weierstrass function $\wp$ satisfies the identity
\begin{equation}
\label{appendix1}
\wp\left( \lambda z; \lambda^{-4}g_2, \lambda^{-6}g_3 \right) = \lambda^{-2} \wp (z; g_2, g_3),
\end{equation}
which holds for any constant $\lambda \neq 0$.

\end{document}